\documentclass[reprint,amsmath,amssymb,aps,pre]{revtex4-1}
\usepackage[utf8]{inputenc}
\usepackage{graphicx}
\usepackage{dcolumn}
\usepackage{mathrsfs,amsmath}
\usepackage{bm}
\usepackage{xcolor}
\usepackage{longtable}
\usepackage{url}
\usepackage{ulem}

\begin{document}
\title{Coarse-graining and criticality in the human connectome}
\author{Youssef Kora and Christoph Simon}
\affiliation{Department of Physics and Astronomy, University of Calgary, Calgary, Alberta, T2N 1N4, Canada}
\affiliation{Hotchkiss Brain Institute, University of Calgary,  Calgary,  Canada}
\date{\today}

\begin{abstract}
 In the face of the stupefying complexity of the human brain, network analysis is a most useful tool that allows one to greatly simplify the problem, typically by approximating the billions of neurons comprising the brain by means of a coarse-grained picture with a practicable number of nodes. But even such relatively small and coarse networks, such as the human connectome with its 100-1000 nodes, may present challenges for some computationally demanding analyses that are incapable of handling networks with more than a handful of nodes. With such applications in mind, we set out to further coarse-grain the human connectome by taking a modularity-based approach, the goal being to produce a network of a relatively small number of modules. We applied this approach to study critical phenomena in the brain; we formulated a hypothesis based on the coarse-grained networks in the context of criticality in the Wilson-Cowan and Ising models, and we verified the hypothesis, which connected a transition value of the former with the critical temperature of the latter, using the original networks. We found that the qualitative behavior of the coarse-grained networks reflected that of the original networks, albeit to a less pronounced extent. This, in principle, allows for the drawing of similar qualitative conclusions by analysing the smaller networks, which opens the door for studying the human connectome in contexts typically regarded as computationally intractable, such Integrated Information Theory and quantum models of the human brain. 
\end{abstract}
\maketitle

\section{Introduction}
The sheer complexity of the human brain constitutes a formidable obstacle to the construction of a complete theory of its workings. Network analysis \cite{sporns_2003} serves as an exceedingly powerful tool that breaks down the problem by modeling the brain, with its billions of neurons, as a network comprising a much more manageable number of nodes. In this simplified picture, it becomes computationally feasible to conduct graph-theoretical  investigations \cite{zalesky_fornito_bullmore_2010,sporns_2017, bullmore_2009} from which a great deal of neuroscientific insight may be extracted. \\ \indent
One such example is the structural human connectome, which is a comprehensive map of neural connections in the brain that may be constructed at different degrees of granularity. It typically comprises a number of nodes of $\sim100$, with some other atlases containing as many as a few hundred \cite{rosen_halgren_2021}. There, each node represents a large collection of neurons, such as an anatomical brain region. The structural connectivity matrix of such a network may be extracted from data obtained through non-invasive neuroimaging techniques such as diffusion tensor imaging \cite{jbabdi_2015,shi_2017,hagmann}.  Patterns of functional activity may also be observed through functional magnetic resonance imaging (fMRI) scans, enabling the construction of the functional human connectome, in which the edges between nodes represent statistical correlations based on similarity measures between neuronal components \cite{raichle, hagmann,lang_tome_keck_2012}. These functional networks may serve as graphical representations of the dynamical patterns that emerge spontaneously in the brain, as well as those which manifest themselves during the performance of tasks \cite{eguiluz_2005, heuvel_2008}. \\ \indent
But even such relatively small and coarse networks are intractable in the context of certain demanding approaches where, for instance, computational costs grow exponentially or super-exponentially with the number of nodes. One such example is the integrated information theory (IIT) of consciousness, which is a framework that seeks to quantify consciousness by introducing the quantity $\Phi$, a number that characterizes the extent to which a dynamical system generates information that is irreducible to the sum of its parts (see Ref. \onlinecite{oizumi} for a comprehensive review). IIT is often regarded as an alternative hypothesis to the other prominent theory of consciousness, global workspace theory \cite{baars_2005}. The cost of computing $\Phi$ explodes combinatorially as a function of the number of nodes, rendering the problem insurmountable for systems comprising more than a handful of nodes.   \\ \indent Furthermore, recent years have seen a rising interest in quantum effects in the brain \cite{adams_2020,smith_2021,hadi_2021}, and in theories which attempt to make connections between quantum mechanics and consciousness \cite{hameroff, fisher}.  Consequently, models of the brain which can incorporate quantum phenomena are of ever-growing interest. Simulations of such quantum models of the brain on a classical computer, however, are computationally expensive in a manner which grows exponentially with the number of nodes. Therefore, to grapple with either of these applications (IIT and quantum models of the brain) in the context of the human connectome, it is essential to employ a coarse-graining approach to reduce the size of the human connectome by at least an order of magnitude.  \\ \indent
To that end we set out to identify a suitable coarse-graining procedure. There is a number of approaches proposed in the literature for detecting communities in networks, such as ones based on centrality indices \cite{girvan_2002} and spectral methods \cite{capocci_2005}. We employ the algorithm proposed in \cite{blondel_2008}, which is a simple and efficient algorithm to partition the networks into communities by maximizing the modularity of the network. \\ \indent
After applying this approach to the human connectome, we investigate the extent to which the resulting coarse-grained networks resemble the original networks. As a sanity check, we start by verifying that the global structures of the simplified networks resemble those of the original ones. Next, we turn our attention to models of the dynamical behavior of the brain, and study the preservation of such behavior in going from the large networks to the small ones. \\ \indent
The first model we consider is that of Wilson-Cowan oscillators \cite{wilsoncowan}, a biologically motivated description of the dynamics of neuronal populations. In the form of choice, the model contains but one free parameter, referred to as $c_5$, which may be varied to control the dynamical state of the network. A salient feature of this model is the presence of a critical point; upon exceeding a certain value of $c_5$, typically denoted as $c_5^T$, the system transitions from a globally inactive state into a globally excited state in which a preponderance of oscillatory activity manifests itself in a great number of brain regions. $c_5^T$, which may be regarded as a given brain network's capacity for global excitation, has been observed to vary across individual subjects \cite{muldoon_2016} and to correlate with a number of cognitive measures \cite{bansal_datadriven}. It has also been demonstrated that biological networks tend to be more easily excitable than randomized networks, despite being less strongly connected \cite{kora_2023}. \\ \indent
Another popular framework that is utilized in this context is the Ising model, by virtue of being one of the simplest models with site-dependent binary variables, and only one parameter, i.e., the temperature of the thermal bath. In one early study \cite{fraiman}, it was found that brain networks derived from the fMRI BOLD (blood-oxygen-level dependent) signals were statistically equivalent to networks derived from the 2D Ising model at the critical temperature. In another \cite{marinazzo}, the Ising model was studied with variable couplings that were defined according to the anatomical connectivity matrix, and total information transfer was found to be maximized at the critical temperature. Both of these investigations lend support to the so-called critical brain hypothesis \cite{bak,massobrio,korchinski_2021}, which conjectures that brain dynamics take place at the so-called edge of chaos, i.e., the critical boundary between stability and disorder. 
 \\ \indent
An Ising model with variable couplings between the spin-sites is often referred to as a generalized Ising model (GIM). Such models have been the subject of a number of recent studies that seek to model the spontaneous activity of the brain given the underlying anatomical structure. In Ref. \onlinecite{abeyasinghe18}, for instance, the couplings in the GIM were defined to be proportional to the number of fibres connecting each pair of brain regions, and the thermodynamics and critical properties of this model were calculated in a Monte Carlo (MC) simulation and compared to those of the 2D Ising model. Correlation functions may also be computed by defining the notion of distance as the reciprocal of the connectivity. A similar study \cite{abeyasinghe20} was performed using different sets of anatomical structural data, including those obtained from patients with severe brain injuries and disorders of consciousness (DOC). This enabled the investigation of the relationship between the critical properties of the model and consciousness, and it was observed that subjects suffering from DOC tended to exhibit higher critical temperatures. \\ \indent
In this work, we evaluate the soundness of the coarse-graining approach using the Wilson-Cowan model by applying it to both the coarse-grained and original networks, and measuring the extent to which the behavior is preserved between them. After verifying that the small networks are adequately representative of the large ones in this dynamical context, we seek to make connections between the transition that takes place  in the in the model of Wilson-Cowan oscillators and criticality in the Ising model. This allows us to hypothesize, based on the coarse-grained networks, a relationship between the transition value of the Wilson-Cowan model $c_5^T$ and the critical temperature of the Ising model $T_c$, thereby introducing the former to the aforementioned association with DOC that was previously observed for the latter (in Ref. \onlinecite{abeyasinghe20}). We finally verify that hypothesis by the aid of the original networks. \\ \indent 
The remainder of this paper is organized as follows: in section \ref{meth} we describe our dynamical models and methodology for data extraction and coarse-graining. We present and discuss our results in section \ref{res}, and finally outline our conclusions in section \ref{conc}.
\section{Materials and Methods}\label{meth}
\subsection{Wilson-Cowan model}\label{wcmo}
We are interested in the form of the Wilson-Cowan model used in Ref. \cite{bansal_datadriven} (but in the absence of an external stimulation), wherein all neurological processes of interest are assumed to be governed by the interaction between excitatory and inhibitory cells. Furthermore, each subpopulation of such cells at every brain region is modeled by a single variable; we define $E_i$ and $I_i$ as the respective fractions of excitatory and inhibitory cells firing per unit time in region $i$. Thus the model reads 
\begin{multline}\label{wilcowe}
\tau \frac{dE_i}{dt} = -E_i(t)+(S_{E_m}-E_i(t)) \times \\ 
S_E \left( c_1E_i(t)-c_2I_i(t) +c_5\sum_j J_{ij}E_j(t-\tau_d^{ij}) \right) \\
+ \sigma w_i(t)
\end{multline}
and
\begin{multline}\label{wilcowr}
\tau \frac{dI_i}{dt} = -I_i(t)+(S_{I_m}-I_i(t))  \times \\ 
S_I \left( c_3E_i(t)-c_4I_i(t) +c_6\sum_j J_{ij}I_j(t-\tau_d^{ij}) \right) \\
+ \sigma v_i(t)
\end{multline}
where $S_{E,I}(x)$ are sigmoid functions given by
\begin{eqnarray}\label{sig}
S_{E,I}(x) = \frac{1}{1+e^{-a_{E,I}(x-\theta_{E,I})}}
- \frac{1}{1+e^{a_{E,I}\theta_{E,I}}}
\end{eqnarray}
$S_{E_m,I_m}$ are the maxima thereof, and the constants $a_{E,I}$ and $\theta_{E,I}$ respectively determine the value and position of maximum slope. $J_{ij}$ are the elements of the structural connectivity matrices. On account of the physical distance $d_{ij}$ between two brain regions, there exists a communication delay $\tau_d^{ij}=d_{ij}/v_d$, where $v_d=10$ m/s is a typical estimate of the signal transmission velocity (see, for instance, Ref. \onlinecite{waxman1972}). Finally, a normal distribution of noise of strength $\sigma$ is injected into the system by means of the functions $w_i(t)$ and $v_i(t)$. This form of the model is standard in the literature \cite{muldoon_2016,bansal_datadriven, bansal_personalized}, but in principle the model may be generalized to incorporate, for instance, couplings between opposing types of cells. 
\\ \indent
We utilized a second order Runge-Kutta solver with a sufficiently fine timestep, such that the results were independent of the size thereof. For each individual subject, the value of $c_5^T$ was estimated by simulating the model at multiple values of $c_5$ and observing the point at which the transition took place.
\\ \indent
\begin{figure}[h]
\centering
\includegraphics[width=0.47\textwidth]{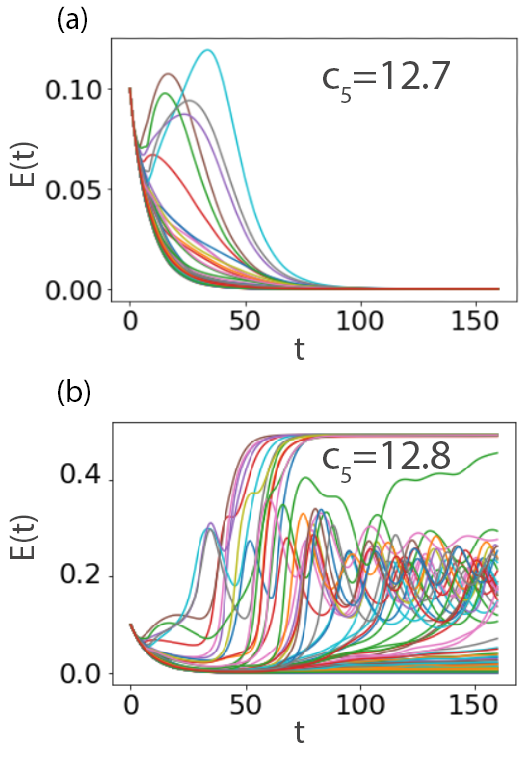}
\caption{An example of the Wilson-Cowan transition: the dynamics of the proportion of excitatory cells firing per unit time, for a certain brain network, at $c_5=12.7$ (a) and $c_5=12.8$ (b). Each line corresponds to a brain region.}
\label{s111211trans}     
\end{figure}
The oscillators were always initialized in the state with $E=I=0.1$. The parameters defined above were fixed at $\sigma=10^{-5}$, $c_1=16$, $c_2=12$, $c_3=15$, $c_4=3$, $a_E=1.3$, $a_I=2$, $\theta_E=4$, $\theta_I=3.7$, and $\tau=8$ as prescribed in the literature \cite{muldoon_2016,bansal_datadriven}. At this choice of the parameters, the oscillators may be found in one of three states: a low fixed point, a high fixed point, and an oscillatory limit cycle in-between \cite{muldoon_2016}. A biological network typically exhibits an abrupt transition into a globally excited state, i.e., from a state where all initial activity rapidly dwindles down to the lower fixed point, to one where a substantial proportion of oscillators are activated into the limit cycle or the high fixed point. Such a transition may be accomplished by ramping up the global coupling parameter $c_5$, and crossing a critical value thereof, at which the state of the system abruptly changes, as illustrated in Fig. \ref{s111211trans}. The transition value, typically referred to as $c_5^T$, will henceforth be referred to as simply $c_5$ for ease of notation. As mentioned above, this value is unique for each subject for a given choice of parameters and initial conditions \cite{analysis}.
\subsection{Generalized Ising model}\label{imo}
\begin{figure}[h]
\centering
\includegraphics[width=0.47\textwidth]{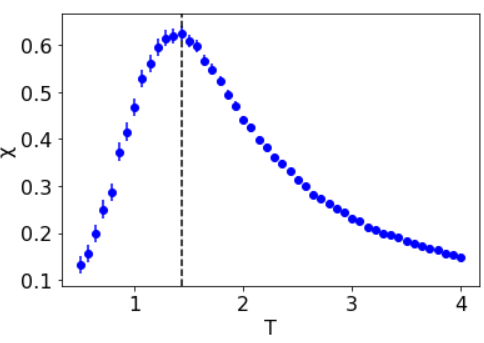}
\caption{An example of the magnetic susceptibility as a function of temperature for a given network of Ising spins. The critical temperature is estimated by locating the peak (black dashed line). Bars correspond to statistical errors, which may be used to infer the uncertainty on the critical temperature by identifying the range of temperature at which the statistical errors overlap with the maximum.}
\label{chi}     
\end{figure}
We consider a system comprising $N$ lattice sites, each with spin $s_i = \pm 1$ along the z-direction. In the absence of an external magnetic field, the energy of a configuration $s = \{s_i\}$ is given by the pairwise interaction
\begin{eqnarray}\label{gimc}
 H (s) = - \sum_{i<j} J_{ij} s_i s_j
\end{eqnarray}
where the sum runs over all pairs of particles. This system is more general than the Ising model, in the sense that a) the sum is  not restricted to nearest neighbours, and b) the coupling strength $J_{ij}$ of the interaction between two lattice sites is arbitrarily non-uniform. Thus it is referred to as the generalized Ising model. 
\\ \indent
Clearly, the physics of this model at any given temperature is completely determined by the coupling strength matrix $\bf J$. which may be chosen to describe an arbitrary model for the interaction between an arbitrary number of spin sites in an arbitrary number of dimensions. 
\\ \indent
As alluded to above, the GIM may serve as a dynamical model of the human brain through the following mapping: each spin site represents a brain region, and the corresponding value of $s_i$ represents the value of a binary variable associated with that brain region, such as the BOLD signal. The coupling strength matrix $J_{ij}$ is mapped onto the structural connectome by, for instance, being made proportional to the number of fibres connecting two regions. 
\\ \indent
We perform simulations of the system described by eq. \ref{gimc}, at each temperature, by means of the Metropolis Monte Carlo algorithm \cite{metropolis} that is the same regardless of the value of $\bf J$. The details of the simulation are standard; we start from a random spin configuration, and sample new configurations with a probability proportional to the corresponding Boltzmann weight, exp$(-E(\{s_i\})/T)$. First, a sufficient number of MC steps is performed to ensure convergence towards thermodynamic equilibrium (a single MC step is defined to be a number of iterations such that, on average, each of the spin sites is subject to one flipping attempt). Once that is achieved, thermal expectation values of the quantities of interest are computed as statistical averages over configurations separated by an appropriate number of MC steps. The magnetization, for instance, is given by
\begin{eqnarray}\label{mag}
m = \frac{1}{N} \sum_{i=1}^N  s_i
\end{eqnarray}
\\ \indent
and the magnetic susceptibility, defined as $\frac{\partial m}{\partial h}\Bigr\rvert_{h \to 0}$, is given by
\begin{eqnarray}\label{sus}
\chi = \frac{N}{T} \left( \langle m^2 \rangle - \langle m \rangle^2 \right) 
\end{eqnarray}
\\ \indent
and the specific heat by
\begin{eqnarray}\label{spec}
C = \frac{N}{T^2} \left( \langle e^2 \rangle - \langle e \rangle^2 \right) 
\end{eqnarray}
where $e=E/N$, the total energy of the system per particle. One may study the phase transition by simulating the system at different temperatures, plotting those quantities thereagainst, and inferring the critical temperature by identifying the peak in the magnetic susceptibility (see Fig. \ref{chi}).  Naturally, the caveats here are 1) that the smaller the network, the more prominent the finite-size effects; and 2) the existence of other methods to estimate the critical temperature.
\subsection{Structural data and measures}\label{strdata}
We constrained the dynamical models outlined in subsections \ref{wcmo} and \ref{imo} by a set of undirected structural connectivity matrices extracted from imaging data, which we obtained from the 1200 subject cohort of the Human Connectome Project (HCP)\href{http://www.humanconnectomeproject.org/}, a database containing neural data for thousands of subjects, of which we selected a sample for our calculations. On no particular basis did we choose from among the 1200 subjects; we merely selected the first subjects on the list, about whom no information was provided besides age and gender. Both pre-processed T1-weighted structural images and 3T dMRI images were used in our computational fibre tracking method. Python DIPY and NiBabel libraries were utilized to perform the streamline calculations using a constrained spherical deconvolution model and probabilistic fibre tracking functions, which are built in the libraries. 
\\ \indent
Each structural connectivity network emerging from this procedure belongs to an individual subject, and comprises 104 nodes, each node corresponding to a cortical or subcortical brain structure. A full list of those brain structures may be found in Table \ref{tab:extended_list} in the Appendix. The networks were then normalized by dividing the strength of each connection by the sum of the
volumes of its two nodes, as in earlier studies \cite{muldoon_2016,bansal_datadriven}. \\ \indent
As our first measure of global structure we calculate the global efficiency \cite{stam_2007}, which measures the extent to which a graph is well-connected by computing the average inverse shortest distance between pairs of vertices. The expression reads
\begin{eqnarray}\label{eglob}
E_{glob} = \frac{1}{N(N-1)}\sum_{i \neq j} \frac{1}{s_{ij}}
\end{eqnarray}
where $N$ is the number of nodes, and $s_{ij}$ is the shortest distance between nodes $i$ and $j$ (not to be confused with the physical distance $d_{ij}$). We also compute the characteristic path length \cite{stam_2007}, which is another (roughly anti-proportional) way of characterizing the well-connectedness of a graph, previously used \cite{kora_2023} in the context of the Wilson-Cowan model \cite{globcharpat}. It is computed as the average path length between all possible pairs of vertices:
\begin{eqnarray}\label{path}
L = \frac{1}{N(N-1)}\sum_{i,j\in N, i \neq j} s_{ij}
\end{eqnarray}

\subsection{Coarse-graining algorithm}\label{coalg}
We employed a coarse-graining algorithm based on the method prescribed in \cite{blondel_2008} for detecting communities in networks. The goal is to partition a given network into the set of communities that maximizes the quantity known as the modularity, which quantifies the strength of the connections \emph{within} communities relative to the strength of the connections \emph{between} communities. In the cases of a weighted network with a connectivity matrix ${\bf J}$, this may be written as
\begin{eqnarray}\label{mod}
Q = \frac{1}{2m}\sum_{i,j}\left[ J_{ij} - \frac{d_i d_j}{2m}  \right] \delta(c_i,cj)
\end{eqnarray}
where $d_i$ is the degree centrality of node $i$, defined as $d_i = \sum_{i=1}^N  J_{ij}$; $c_i$ is the community to which node $i$ belongs; $\delta(u,v)$ is a Kronecker delta, equaling 1 if u=v and 0 otherwise; and $m=\frac{1}{2}\sum_{ij}J_{ij}$. \\ \indent
The algorithm comprises a series of passes, each pass consisting of two phases. Phase one begins in a maximum entropy configuration with each node assigned to its own community, such that the number of communities is equal to the number of nodes. Next, each neighbouring node $j$ of node $i$ is considered, and we compute the gain in modularity on account of moving node $i$ into the community containing node $j$. Node $i$ is subsequently removed from the community to which it originally belonged, and placed into the community for which the gain in modularity is maximum (provided that gain is positive). All nodes in the network are subjected to this process sequentially and repeatedly, until the modularity may no longer be increased by any such moves, thereby reaching a local maximum and concluding phase one. \\ \indent
To evaluate the change in modularity arising from the moving of an isolated node $i$ into a community $C$, we make use of the computationally efficient expression given in Ref. \cite{blondel_2008}, namely
\begin{multline}\label{delmod}
\Delta Q = \left[ \frac{\Sigma_{in} + k_{i,in}}{2m} - \left(\frac{\Sigma_{tot}+k_i}{2m} \right)^2 \right] - \\ \left[ \frac{\Sigma_{in}}{2m} - \left(\frac{\Sigma_{tot}}{2m}\right)^2 -
\left(\frac{k_i}{2m}\right)^2 \right]
\end{multline}
where $\Sigma_{in}$ is the sum of the weights within $C$, $\Sigma_{tot}$ is the sum of the weights incident to nodes in $C$, and $k_{i,in}$ is the sum of the weights from $i$ to nodes in $C$. \\ \indent
Phase two of the algorithm is concerned with the construction of a new network whose nodes are the communities obtained in phase one. In this new network, the link connecting two nodes is obtained by summing up the weights of the links between the the two communities which the two nodes represent. This obviously gives rise to self-loops, which may either be represented as self-interactions or ignored altogether. For the purposes of this work, we do not incorporate self-interactions, as the original networks did not contain anything of the sort. It is important to note that the weights of the non-self loop links are not renormalized at any step of the algorithm. We also compute the distances between nodes in the new network. Now that the nodes are no longer strictly correspondent to anatomical brain regions, a new definition of the distance is in order; we define the distance between two communities as the weighted mean of the distances between the nodes therein.   \\ \indent
We arrive at the partitioning which maximizes the modularity by iterating the passes of the algorithm until no more increases to the modularity may be achieved. \\ \indent
\section{Results and Discussion}\label{res}
\begin{figure}[h]
\centering
\includegraphics[width=0.47\textwidth]{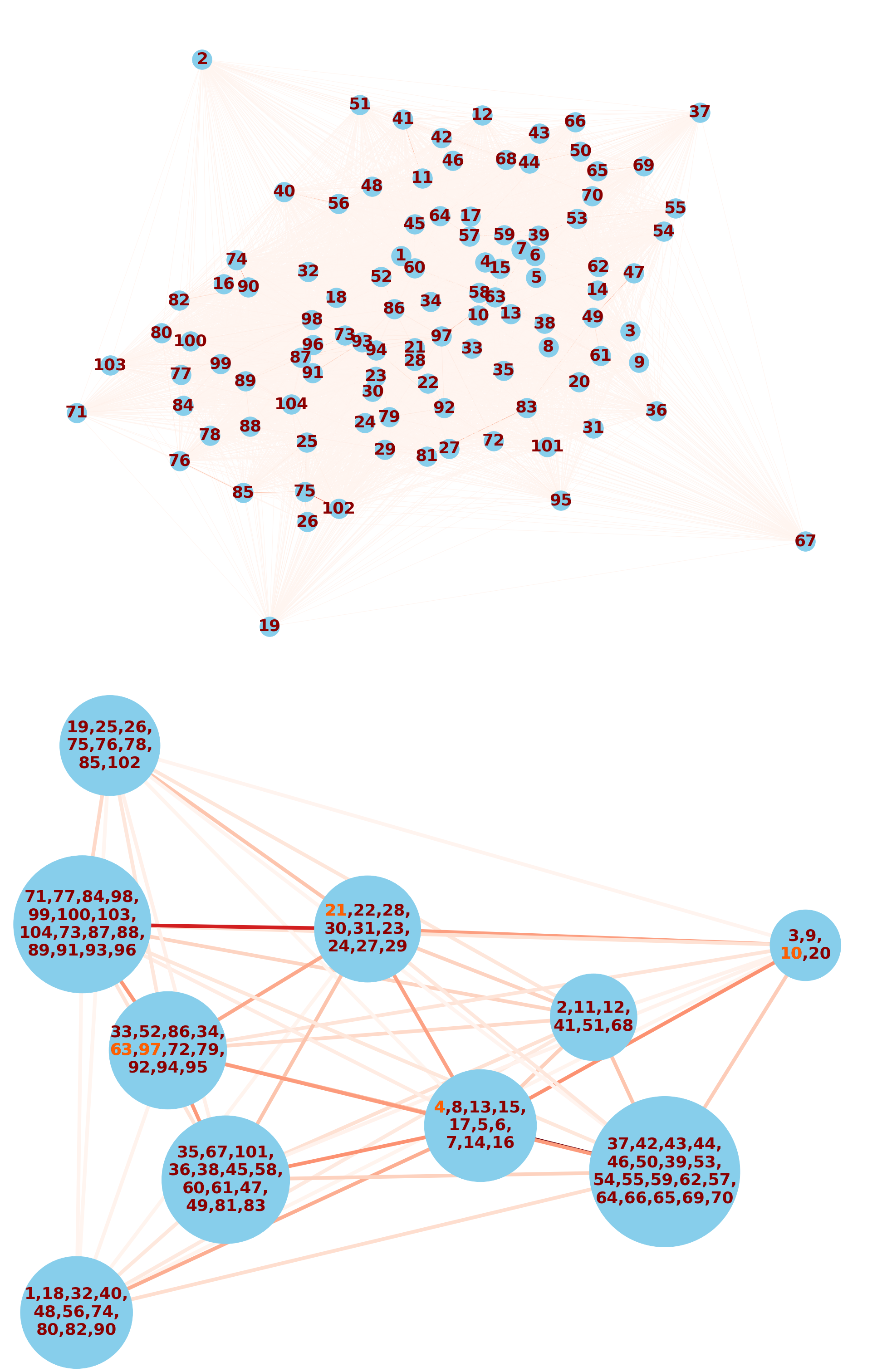}
    \caption{A visualization of the relationship between one brain network (top) and its coarse-grained counterpart (bottom). Numbers correspond to brain regions in accordance with the list given in Table \ref{tab:extended_list} in the Appendix. The darkness of a given edge is proportional to its weight. The area of a given node in the coarse-grained network is proportional to the number of brain regions it contains. Highlighted numbers in the coarse-grained network correspond to the most salient brain regions according to Ref. \cite{salhi_2022}}.
\label{visual}     
\end{figure}
\begin{figure}[h]
\centering
\includegraphics[width=0.47\textwidth]{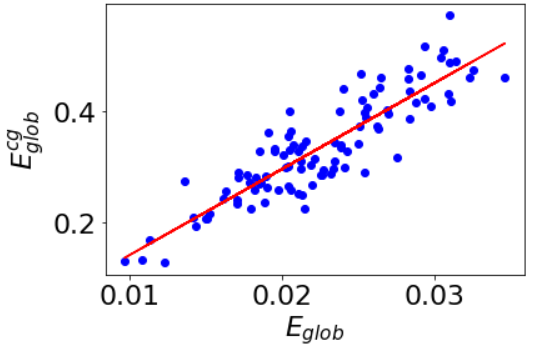}
    \caption{The global efficiency of the coarse-grained networks obtained vs. the global efficiency of the corresponding original networks. Each dot represents a different subject. The red line is a least-squares fit with $r^2=0.786$.}
\label{globglob}     
\end{figure}
The outcome of applying the procedure outlined in subsection \ref{coalg} to the networks described in subsection \ref{strdata} is a reduction of any given 104-node network to one comprising 10-14 nodes, depending on the subject. One such network and its corresponding coarse-grained network are visualized in Fig. \ref{visual}. Each number corresponds to a brain region as given by the list in Table \ref{tab:extended_list}. The most prominent brain regions, as reported by the centrality analyses conducted in Ref. \cite{salhi_2022}, include the brainstem (10), and both sides of the superiorfrontal cortex (63 and 97) and thalamus (4 and 21), all of which regions are highlighted in orange. \\ \indent
We now investigate the degree to which these small networks are representative of the original networks from which they emerged. Henceforth, we attach to all variables relating to the maximally coarse-grained networks the superscript ``$cg$", e.g., $T_c^{cg}$ is the critical temperature of the coarse-grained networks.  \\ \indent
We begin by looking at global structure. We computed the global efficiency, as defined in subsection \ref{strdata}, for each subject's network and the corresponding maximally coarse-grained network. The result of which computation is presented in Fig. \ref{globglob}, which shows a fairly strong ($r^2=0.879$) correlation between the global efficiencies of both types of network. This is an encouraging result, as it indicates that the original networks of 104 nodes are fairly well-represented by the networks of 10-14 nodes which emerged from our coarse-graining procedure. 
\begin{figure}[h]
\centering
\includegraphics[width=0.47\textwidth]{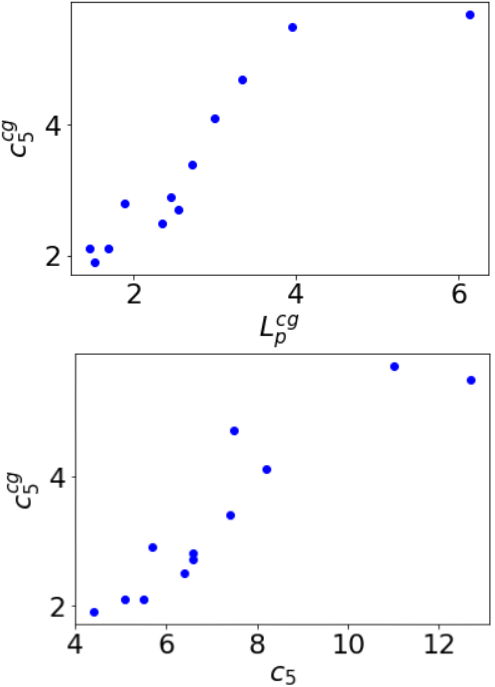}
    \caption{The transition value of $c_5$ for the coarse-grained networks against the (top) corresponding characteristic path length ($r^2=0.828$) and (bottom) transition value of $c_5$ for the corresponding original network ($r^2=0.859$).}
\label{c5ult}     
\end{figure}
\\ \indent Emboldened by this finding, we turn our attention to dynamical behavior. We start by asking the question of whether we may arrive at similar conclusions to those of Ref. \onlinecite{kora_2023}, in which the full-sized networks were studied, by instead looking at the coarse-grained networks. We do so by implementing the Wilson-Cowan model on our small networks. \\ \indent
The first relationship we expect to see, based on the results of Ref. \onlinecite{kora_2023}, is a positive correlation between the transition value and the corresponding characteristic path length. This is an observation that the more well-connected a biological network is, the more global excitable it tends to be. Fig. \ref{c5ult} (top) shows the result of this investigation, which clearly assures us that this behavior is enjoyed by the coarse-grained networks as it is by the original ones. We also see in \ref{c5ult} (bottom) a substantial positive correlation $r^2=0.859$ between the the transition value for the small networks and the large ones, further validating the similarity between the two types of networks. \\ \indent
It must be noted here that such strong correlations are contingent on performing the coarse-graining procedure in the form prescribed in subsection \ref{coalg}. We have experimented with some deviations from that prescription, which have invariably degraded  the extent to which the behavior of the coarse-grained networks represented that of the original ones. For instance, this representation was slightly diminished when we tried taking the connectivity between modules to be the average of edge weights rather than their sum, as we show in Fig. \ref{averaged} in the Appendix. The representation was also greatly diminished when we attempted to normalize the weights of the non-self loop links in the intermediate steps of the algorithm (to wit by dividing the weights of the links by the weight of the greatest link among them after each pass of the algorithm); the correlation between $c_5^{cg}$ and $c_5$ was completely lost, as well as that between the $E_{glob}^{cg}$ and $E_{glob}$. This breakdown suggests that the normalizing at intermediate steps leads to the erasure of important information as to the weights of the inter-module links with respect to the original networks. Indeed, when the prescription outlined above is adhered to, the behavior is consistent across multiple coarse-graining scales, as shown in Fig. \ref{intermediate} in the Appendix.  \\ \indent
\begin{figure}[h]
\centering
\includegraphics[width=0.47\textwidth]{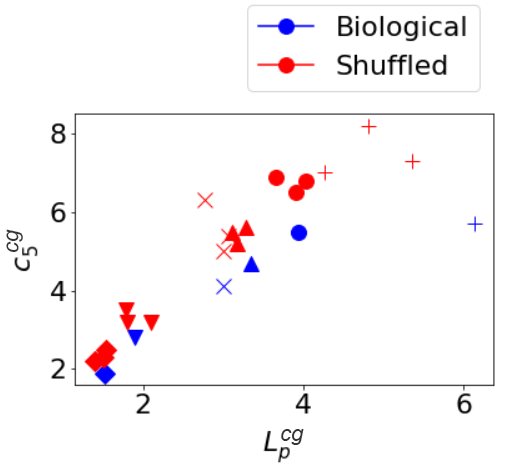}
    \caption{The transition value of $c_5$ for the coarse-grained networks against the corresponding characteristic path length, for both biological networks (blue) and shuffled networks (red). Symbol shapes corresponds to indivudal subjects.}
\label{withshuffled}     
\end{figure}
The second conclusion drawn in Ref. \onlinecite{kora_2023} is that randomized networks tend to be less globally excitable (i.e., higher value of $c_5$) than biological ones, despite being more well-connected (lower characteristic path length). This phenomenon is also observed for coarse-grained networks networks, albeit substantially less pronounced, as indicated in Fig. \ref{withshuffled}. We can clearly see that the red line (shuffled networks, generated by shuffling pairwise connectivities while preserving their distribution) invariably lies above the blue line (biological networks of the same subjects), but by a significantly smaller amount compared to the case of the original networks, as may be seen in Ref. \onlinecite{kora_2023}. This suggests that by studying the coarse-grained networks in lieu of the full-sized ones, one may derive qualitatively similar conclusions, but might fail to observe the full quantitative extent thereof. It may be argued that the red line is closer to the blue line in the coarse-grained networks, compared to the original ones, on account of the smaller effect of shuffling in smaller networks with respect to larger ones. \\ \indent
At this stage, it's clear that the reduced networks produced from the coarse-graining procedure retain not only significant structural properties, but also a good deal of dynamical behavior. Thus we proceed to investigate the Ising model for the coarse-grained system, in the hopes that the conclusions we draw represent reality nearly as much as those directly drawn from the human connectome itself. \\ \indent
\begin{figure}[h]
\centering
\includegraphics[width=0.47\textwidth]{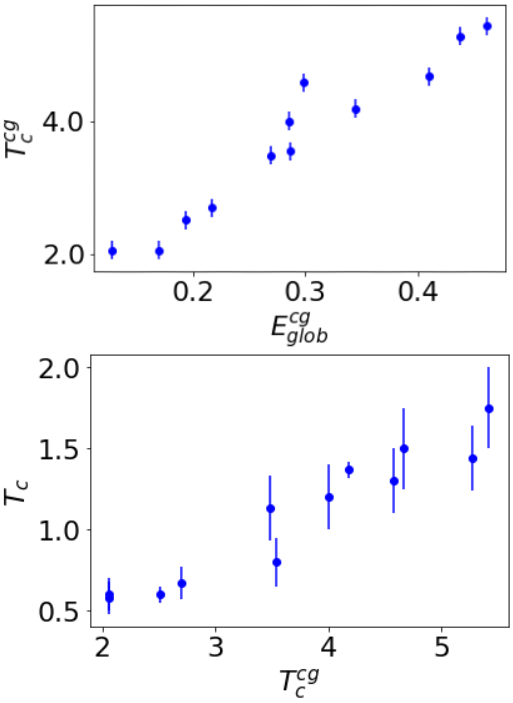}
    \caption{The Ising critical temperature as a function of global efficiency, for the coarse-grained networks ($r^2=0.930$) (top), and the correlation between the critical temperatures in the full and coarse-grained pictures ($r^2=0.917$) (bottom).}
\label{Tccg}     
\end{figure}
We begin by computing the Ising critical temperature, as described in subsection \ref{imo} for a variety of coarse-grained networks, and investigating how it varies with the global efficiency introduced in subsection \ref{strdata}. The advantage of simulating the Ising model on such small systems is that the simulations are rapid and efficient, and so the statistical errors of the thermal expectation values may be shrunk arbitrarily, allowing for an arbitrarily precise estimation of the critical temperature. The result of this is displayed in Fig. \ref{Tccg} (top), which shows a strong positive correlation ($r^2=0.930$), indicating that well-connected networks possess higher critical temperatures. As a sanity check, we compare the coarse-grained critical temperatures with those computed for the original networks in Fig. \ref{Tccg} (bottom), in order to verify that there is indeed a strong correlation, notwithstanding the size of the uncertainties on the critical temperature; they are inferred from the magnetic susceptibility data by estimating the range in which the peak is expected to lie in light of the statistical errors. In the case of the original networks, those statistical errors are substantial, they being the product of rather ponderous simulations in which the accumulation of statistics takes a long time.\\ \indent
\begin{figure}[h]
\centering
\includegraphics[width=0.47\textwidth]{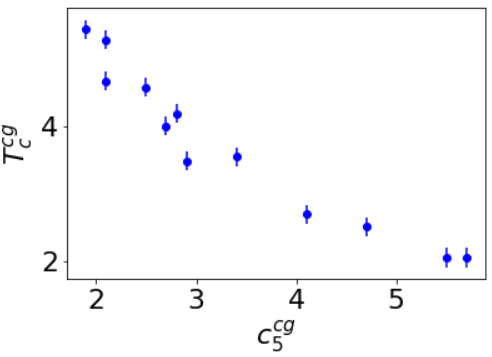}
    \caption{The Ising critical temperature as a function of the Wilson-Cowan transition value, for the coarse-grained networks ($r^2=0.918$)}.
\label{Tcultc5ult}     
\end{figure}
Now that we have established that the behavior of the coarse-grained networks is fairly well-representative of the original ones, we are poised to use the coarse-grained networks to connect the two dynamical models and the two different senses of criticality therein. In Fig. \ref{Tcultc5ult}, we plot the Ising critical temperature for the coarse-grained networks, $T_c^{cg}$, against the Wilson-Cowan transition value, $c_5^{cg}$. We clearly see a negative correlation between the two quantities, which is consistent with the findings presented in  both Figs. \ref{c5ult} and \ref{Tccg}, taken together with the fact that the characteristic path length and the global efficiency are opposing measures. The inverse relationship between $T_c$ and $c_5$ illustrated in Fig. \ref{Tcultc5ult} suggests that highly excitable brain networks (i.e., with low values of $c_5^T$, denoted as $c_5$), which is a property enjoyed by biological networks relative to randomized ones \cite{kora_2023}, tend to possess higher critical temperatures. This here is a surprising result when considered with reference to one of the findings of Ref. \onlinecite{abeyasinghe20}, namely the association of a high Ising critical temperature with the presence of a disorder of consciousness. Naively, one might have expected randomized networks to  exhibit behavior closer to DOC networks than to healthy ones, but here we have a set of observations that seem to suggest that when it comes to a network's global excitability, shuffled networks possess the least, followed by biological networks (which observation may be directly seen in Fig. \ref{withshuffled} and is presented and discussed in more detail in Ref. \onlinecite{kora_2023}), followed by the most excitable networks of all: DOC networks (which may be inferred from Fig. \ref{Tcultc5ult} in conjunction with the association of high $T_c$ with DOC reported in Ref. \cite{abeyasinghe20}). \\ \indent
\begin{figure}[h]
\centering
\includegraphics[width=0.47\textwidth]{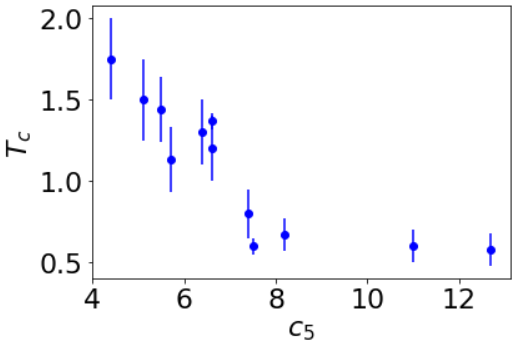}
    \caption{The Ising critical temperature a function of the Wilson-Cowan transition value for the original networks ($r^2=0.682$).}
\label{Tcfull}     
\end{figure}
Finally, we may validate this conclusion drawn from the simplified networks, provided we are prepared to contend with the sluggishness of the Ising simulations on the unsimplified networks, and with substantial uncertainties arising from substantial statistical errors. The results, shown in Fig. \ref{Tcfull}, are certainly encouraging; the relationship between the Ising critical temperature and the Wilson-Cowan transition value bears a good deal of resemblance to the coarse-grained case, notwithstanding the disparity in the sizes of the uncertainties. This lends further credence to the belief that the coarse-graining procedure yields networks that are both significantly simpler to work with and adequately representative of the original networks, and that one may ascribe considerable validity to hypotheses developed based on the coarse-grained networks.

\section{Conclusions}\label{conc}
Inspired by the idea that the human connectome itself is indeed a coarse-grained picture, we set out to effect further coarse-graining in an effort to render the networks small enough for analyses of exceeding computational cost. To that end, we employed a modularization approach that allowed us to reduce the 104-node networks to networks comprising but 10-14 nodes. 
\\ \indent
These coarse-grained networks were found to be adequately representative of the original networks in both structural and dynamical contexts, provided that the coarse-graining procedure is carried out in a certain form. Encouraged by this, we used the small networks to formulate a hypothesis concerning the relationship between two dynamical models: Wilson-Cowan and Ising. Finally, we used the unsimplified networks to verify our hypothesis, which suggests that our approach may be generally viable, and emboldens us to pursue further applications. \\ \indent
Such applications include 
Integrated Information Theory and theories of the quantum brain, where the computational complexity is exponential or super-exponential in the number of nodes. The human connectome in its traditional form typically comprises 100-1000 nodes, which is far too large a number for any such applications. Fortunately, our findings here indicate that conclusions drawn based on coarse-grained networks may be valid enough, which provides a path to more explicit analyses in the aforementioned contexts. 
\section{Acknowledgments}\label{ack}
This work was supported by the Natural Sciences and Engineering Research Council (NSERC) of Canada through its NSERC Discovery Grant Program, the Alberta Major Innovation Fund, National Research Council through its Applied Quantum Computing Challenge Program, and Quantum City. We would also like to thank Joern Davidsen for the useful discussions and feedback, and Salma Salhi for providing the structrual data. Additionally, we thank the HCP for providing access to their data. \\
\section*{Data Availability Statement}
The data used in this project was provided by the Human Connectome Project (HCP; Principal
Investigators: Bruce Rosen, M.D., Ph.D., Arthur W. Toga, Ph.D., Van J. Weeden, MD). HCP funding was provided by the
National Institute of Dental and Craniofacial Research (NIDCR), the National Institute of Mental Health (NIMH), and the
National Institute of Neurological Disorders and Stroke (NINDS). HCP data are disseminated by the Laboratory of Neuro
Imaging at the University of Southern California. Structural and diffusion MRI images from the HCP, as well as lists of extracted structures, bvals, and bvecs, were all used to process the data in our Python program. All subjects are part of the ``WU-Minn HCP Data - 1200 Subjects" \href{https://db.humanconnectome.org/data/projects/HCP_1200}{dataset}. A complete list of subject names is available upon request. All scripts used to generate the connectomes are available on our \href{https://github.com/SalmaSalhi7/Structural-Connectome-Project}{Github} repository found at \url{https://github.com/SalmaSalhi7/Structural-Connectome-Project}.

\bibliography{biblio}

\begin{thebibliography}{40}
\bibitem [{glo()}]{globcharpat}%
  \BibitemOpen
  \href@noop {} {}\bibinfo {note} {While global efficiency and characteristic
  path length are equivalent (and opposing) ways to measure the
  well-connectedness of a graph, we chose to employ the former for the bulk of
  our calculations for computational efficiency reasons. However, where it
  concerns the Wilson-Cowan model, we chose to compute the characteristic path
  length for a sample of networks in order to simplify the process of comparing
  with previous work that utilized the characteristic path length to measure
  well-connectedness.}\BibitemShut {Stop}%
\end{thebibliography}

\section{Appendix}
\subsection{List of Brain regions}

 \begin{longtable}{|c | c |  c| c| c|} 
  \caption{A full list of the 104 brain structures.}
\\ \hline
ID & Structure  
\\ \hline
1 & Left-Lateral-Ventricle\\ 
2 & Left-Inf-Lat-Vent  \\ 
3 & Left-Cerebellum-Cortex  \\
4 & Left-Thalamus-Proper \\ 
        5 & Left-Caudate \\ 
        6 & Left-Putamen  \\ 
        7 & Left-Pallidum \\ 
        8 & 3rd-Ventricle  \\ 
        9 & 4th-Ventricle  \\ 
        10 & Brain-Stem  \\ 
        11 & Left-Hippocampus  \\ 
        12 & Left-Amygdala  \\ 
        13 & CSF  \\ 
        14 & Left-Accumbens-area  \\ 
        15 & Left-VentralDC \\ 
        16 & Left-vessel  \\ 
        17 & Left-choroid-plexus \\ 
        18 & Right-Lateral-Ventricle  \\ 
        19 & Right-Inf-Lat-Vent  \\ 
        20 & Right-Cerebellum-Cortex  \\ 
        21 & Right-Thalamus-Proper  \\ 
        22 & Right-Caudate  \\ 
        23 & Right-Putamen \\ 
        24 & Right-Pallidum \\ 
        25 & Right-Hippocampus  \\ 
        26 & Right-Amygdala \\ 
        27 & Right-Accumbens-area  \\ 
        28 & Right-VentralDC \\ 
        29 & Right-vessel  \\ 
        30 & Right-choroid-plexus  \\ 
        31 & Optic-Chiasm  \\ 
        32 & CC\_Posterior \\ 
        33 & CC\_Mid\_Posterior \\ 
        34 & CC\_Central \\ 
        35 & CC\_Mid\_Anterior \\ 
        36 & CC\_Anterior  \\ 
        37 & ctx-lh-bankssts \\ 
        38 & ctx-lh-caudalanteriorcingulate \\ 
        39 & ctx-lh-caudalmiddlefrontal  \\ 
        40 & ctx-lh-cuneus  \\ 
        41 & ctx-lh-entorhinal  \\ 
        42 & ctx-lh-fusiform  \\ 
        43 & ctx-lh-inferiorparietal  \\ 
        44 & ctx-lh-inferiortemporal  \\ 
        45 & ctx-lh-isthmuscingulate  \\ 
        46 & ctx-lh-lateraloccipital \\ 
        47 & ctx-lh-lateralorbitofrontal  \\ 
        48 & ctx-lh-lingual \\ 
        49 & ctx-lh-medialorbitofrontal \\ 
        50 & ctx-lh-middletemporal  \\ 
        51 & ctx-lh-parahippocampal  \\ 
        52 & ctx-lh-paracentral  \\ 
        53 & ctx-lh-parsopercularis \\ 
        54 & ctx-lh-parsorbitalis  \\ 
        55 & ctx-lh-parstriangularis \\ 
        56 & ctx-lh-pericalcarine  \\ 
        57 & ctx-lh-postcentral \\ 
        58 & ctx-lh-posteriorcingulate \\ 
        59 & ctx-lh-precentral  \\ 
        60 & ctx-lh-precuneus  \\ 
        61 & ctx-lh-rostralanteriorcingulate \\ 
        62 & ctx-lh-rostralmiddlefrontal \\ 
        63 & ctx-lh-superiorfrontal  \\ 
        64 & ctx-lh-superiorparietal \\ 
        65 & ctx-lh-superiortemporal  \\ 
        66 & ctx-lh-supramarginal  \\ 
        67 & ctx-lh-frontalpole  \\ 
        68 & ctx-lh-temporalpole \\ 
        69 & ctx-lh-transversetemporal \\ 
        70 & ctx-lh-insula \\ 
        71 & ctx-rh-bankssts  \\ 
        72 & ctx-rh-caudalanteriorcingulate \\ 
        73 & ctx-rh-caudalmiddlefrontal  \\ 
        74 & ctx-rh-cuneus \\ 
        75 & ctx-rh-entorhinal \\ 
        76 & ctx-rh-fusiform  \\ 
        77 & ctx-rh-inferiorparietal  \\ 
        78 & ctx-rh-inferiortemporal \\ 
        79 & ctx-rh-isthmuscingulate  \\ 
        80 & ctx-rh-lateraloccipital  \\ 
        81 & ctx-rh-lateralorbitofrontal \\ 
        82 & ctx-rh-lingual  \\ 
        83 & ctx-rh-medialorbitofrontal  \\ 
        84 & ctx-rh-middletemporal  \\ 
        85 & ctx-rh-parahippocampal\\ 
        86 & ctx-rh-paracentral \\ 
        87 & ctx-rh-parsopercularis \\ 
        88 & ctx-rh-parsorbitalis \\ 
        89 & ctx-rh-parstriangularis \\ 
        90 & ctx-rh-pericalcarine  \\ 
        91 & ctx-rh-postcentral \\ 
        92 & ctx-rh-posteriorcingulate  \\ 
        93 & ctx-rh-precentral  \\ 
        94 & ctx-rh-precuneus  \\ 
        95 & ctx-rh-rostralanteriorcingulate \\ 
        96 & ctx-rh-rostralmiddlefrontal  \\ 
        97 & ctx-rh-superiorfrontal \\ 
        98 & ctx-rh-superiorparietal  \\ 
        99 & ctx-rh-superiortemporal \\
        100 & ctx-rh-supramarginal  \\ 
        101 & ctx-rh-frontalpole  \\ 
        102 & ctx-rh-temporalpole \\ 
        103 & ctx-rh-transversetemporal  \\ 
        104 & ctx-rh-insula  \\ [1ex] 
 \hline 
 \end{longtable}
        \label{tab:extended_list}

\subsection{Supplementary Figures} 
\subsubsection{Coarse-graining with averaged connectivity}
\begin{figure}[h]
\centering
\includegraphics[width=0.47\textwidth]{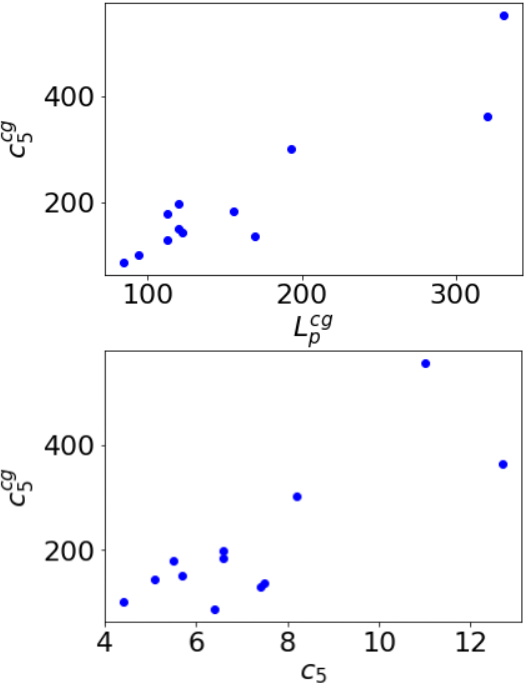}
    \caption{The same as Fig. \ref{c5ult}, but inter-community connectivity is defined by taking the average of node weights rather than the sum. Errors are small than symbol sizes. $r^2=0.854$ and $r^2=0.662$}
\label{averaged}     
\end{figure}
Fig. \ref{averaged} presents the Wilson-Cowan results for simplified networks that are a product of a slightly different coarse-graining procedure; the weights of the links between two communities is defined as the {\em average} (rather than the sum) of the weights of the links connecting the nodes constituting the two communities. This results in a worsening of the correlations, which suggests that using the sum of weights is the more conducive method to our goal of producing coarse-grained networks that are adequately representative of the original ones.
\subsubsection{Intermediate coarse-graining}
Finally, we conducted the same analysis for coarse-grained networks of intermediate sizes, computed using the same algorithm but terminated before the maximum of modularity is attained. The results are clearly unchanged in this third scale, as may be observed in Fig. \ref{averaged}, which is perhaps unsurprising, but it serves as a sanity check. 

\begin{figure}[t]
\centering
\includegraphics[width=0.47\textwidth]{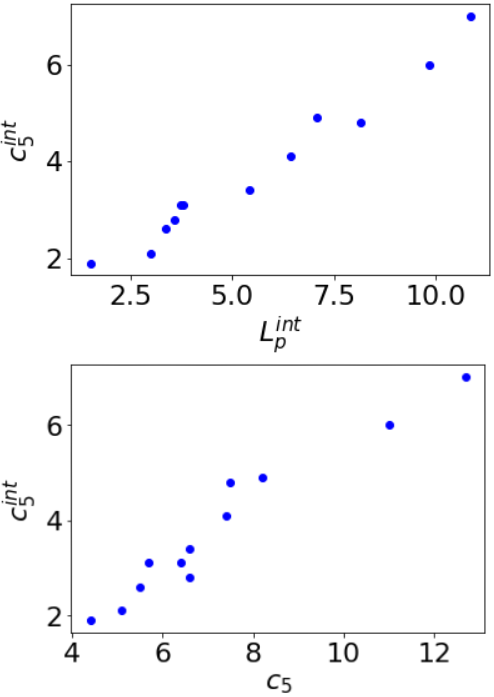}
    \caption{The same as Fig. \ref{c5ult}, but for intermediately coarse-grained networks. Errors are small than symbol sizes. $r^2=0.970$ and $r^2=0.939$.}
\label{intermediate}     
\end{figure}
\end{document}